\documentclass{appolb}
\usepackage{epsfig}
\usepackage{amssymb}

\begin{document}
\title{Formation of $\eta$ Mesic Nuclei%
\thanks{Presented at International Symposium on Mesic Nuclei, Jagiellonian University, Krakow, Poland, June 16, 2010.}%
}
\author{Satoru Hirenzaki and Hideko Nagahiro
\address{Department of Physics, Nara Women's University, Nara 630-8506, Japan}
\and
Daisuke Jido
\address{Yukawa Institute for Theoretical Physics, Kyoto University, Kyoto 606-8502, Japan}
}
\maketitle
\begin{abstract}
In this paper, we briefly introduce the interests and the recent research activities on the $\eta$ mesic nuclei.  We also mention the activities on the  $\eta' (958)$ mesic nucleus formation .  
\end{abstract}
  
\section{Introduction}
Meson-nucleus bound systems are considered to be very interesting objects in the contemporary 
hadron-nuclear physics.  Especially, we can make use of them as $laboratories$ for the studies of hadron properties 
at finite density which are closely related to the symmetry breaking and restoration patters of the strong interaction \cite{Hatsuda}.  

One of the most interesting results was obtained by observation of deeply bound pionic atoms \cite{toki}.  Since the $s$-wave strength of the pion-nucleus optical potential is considered to 
provide the information on the pion decay constant $f_\pi$ in nucleus, it is extremely interesting to study the $s$-wave part of the potential.  For the purpose, it is highly required to observe the pionic $s$-states precisely because the $s$-states properties strongly depend on the $s$-wave potential \cite{ume} and, thus, the $1s$ states in heavy nuclei ($N>Z$) provide key information on the isovector part of the $s$-wave potential, which is related to $f_\pi$ through the Tomozawa-Weinberg theorem \cite{KKWJHK}.  According to this line, Suzuki $et$ $al$. \cite{Suzuki} performed an experiment based on the theoretical predictions \cite{ume,zaki}, and obtained excellent new data of the deeply bound pionic $1s$ states in Sn isotopes,   
which helped much to deduce the new experimental information on the partial restoration of chiral symmetry in nucleus \cite{Suzuki}. 
Thus, it is extremely interesting to extend this research field of meson bound systems to include other mesic systems.    

In this paper, we report the recent research activities on the $\eta$ mesic nuclei.  
We also mention shortly the recent research activities on $\eta ' (958)$ mesic nucleus.

\section{$\eta$ -Nucleus systems}
We discuss the $\eta$-nucleus systems in this report. 
We are interested in the $\eta -$mesic nuclei because they can be considered as a doorway 
to the study of the $N^*(1535)$ resonance $(N^{*})$ in nucleus due to the strong coupling of $\eta N$ and $N^{*}$ \cite{PDG}. The nature of the $N^{*}$ is still controversial and there exist two different pictures of $N^{*}$,  which provide significantly different $\eta -$nucleus potentials as 
shown below.  

Around twenty years ago, the $\eta$-mesic nuclei were studied by Haider and Liu~\cite{HaiderLiu}
and by Chiang, Oset and Liu~\cite{PRC44(91)738}. As for the formation
reaction, the attempt to find the bound states by the ($\pi^+$,p)
reaction led to a negative result~\cite{PRL60(88)2595}. Recently,
some experiments in photoproduction processes 
indicated
observations of such bound
states in $^{12}$C target~\cite{Sokol} and $^3$He target~\cite{Pfeiffer}.  
A result of the proton induced reaction was reported in Ref.~\cite{GEM}.  
New experiments are also planed to 
investigate further 
the $\eta$-nucleus interaction and possible bound states~\cite{Moskal,Itahashi}. 

\subsection{$\eta-$Nucleus Interaction based on $N^*(1535)$ dominance}
\label{sec:etapot}

Here, we briefly explain how the $N^*$ nature affects the $\eta-$nucleus optical potential~\cite{jido02} using two theoretical models.   
In the $\eta$-nucleon system, the $N^{*}$ resonance 
plays an important role due to the 
dominant
$\eta NN^{*}$ coupling.
Here we evaluate
the $\eta$-nucleus optical potential $V_{\eta}(\omega,\rho(r))$
in the two different models 
which are
based on distinct physical pictures of $N^{*}$. One
is the chiral doublet model.
This 
is an extension of the linear sigma model
for the nucleon and its chiral
partner~\cite{PRD39(89)2805,PTP106(01)873etc,PRD57(98)4124}.
We adopted a mirror assignment of the chiral doublet model in this paper.  
The other is the chiral unitary model, in which
$N^{*}$ is dynamically generated in the coupled channel 
meson-baryon scattering~\cite{PLB550,NPA612}.  
We will compare the phenomenological consequences of these models based on the common assumption, the $N^{*}$ dominance of the $\eta$-nucleon system.  

In the first approach, the $N^{*}$ is introduced as a particle with a large
width and appears in an effective Lagrangian together with the nucleon field.
Assuming $N^{*}$-hole excitation induced by the $\eta$ meson in nucleus,
we obtain
the $\eta$-nucleus optical potential at finite nuclear density
as,
\begin{equation}
V_\eta(\omega,\rho(r))
= \frac{g_\eta^2}{2\mu}\frac{\rho(r)}{\omega+m^*_N(\rho)
-m^*_{N^*}(\rho)+i\Gamma_{N^*}(\omega,\rho)/2
},
\label{eq:eta-potential}
\end{equation}
in the local density approximation and the heavy baryon
limit\cite{PRC44(91)738}. Here
$\mu$ is the $\eta$-nucleus reduced mass
and
$\rho(r)$ is the density distribution of the nucleus.
The $\eta NN^{*}$ coupling is assumed to be $S$-wave:
\begin{equation}
 {\cal L}_{\eta NN^*}(x)=g_\eta \bar{N}(x)\eta(x)N^*(x)+{\rm H.c.},
\label{eq:Lagrangian_etaNN}
\end{equation}
and the coupling constant $g_\eta$ is determined to be $g_\eta \simeq 2.0 $
in order to
reproduce the partial width $\Gamma_{N^*\rightarrow\eta N} \simeq 75$MeV
at tree level.
$m^*_N$ and $m^*_{N^*}$ are the effective
masses of $N$ and $N^*$ in the nuclear medium, respectively.
Considering that the $N^{*}$ mass in free space lies only 50 MeV above the
threshold and that the mass difference of $N$ and $N^{*}$ might 
change
in the medium, the $\eta$-nucleus optical potential is expected to be extremely
sensitive to the in-medium properties of $N$ and $N^{*}$. For instance, if the
mass difference reduces in the nuclear medium as $m_{\eta} + m_{N}^{*} -
m_{N^{*}}^{*} > 0$, then the optical potential turns to be repulsive
\cite{jido02}. 

In the chiral doublet model together with the expected partial
restoration of chiral symmetry, a reduction of the mass difference of $N$ and
$N^{*}$ in the medium is found to be 
in the mean field approximation as
\cite{PRD39(89)2805,PTP106(01)873etc,PLB224(89)11,NPA640(98)77},
\begin{equation}
m^*_N(\rho)-m^*_{N^*}(\rho)=\Phi(\rho)(m_N-m_{N^*}),
\label{eq:mass_dif}
\end{equation}
where $m_{N}$ and $m_{N^{*}}$ are the $N$ and $N^{*}$ masses in free space,
respectively, and
\begin{eqnarray}
\Phi(\rho)=1-C{ \rho \over \rho_0} \ . \label{eq:defPhi}
\end{eqnarray}
Here we take the linear density approximation of the in-medium modification of
the chiral condensate, and the parameter $C$ represents the strength of the
chiral restoration at the nuclear saturation density $\rho_{0}$. The empirical
value of $C$ lies from 0.1 to 0.3~\cite{PRL82}. Here we perform our
calculations with
$C=0.0$ and $0.2$ in order to clarify the effects of the partial restoration
of chiral symmetry. 

\begin{figure}[hbt]
\epsfxsize=11cm
\centerline{
\epsfbox{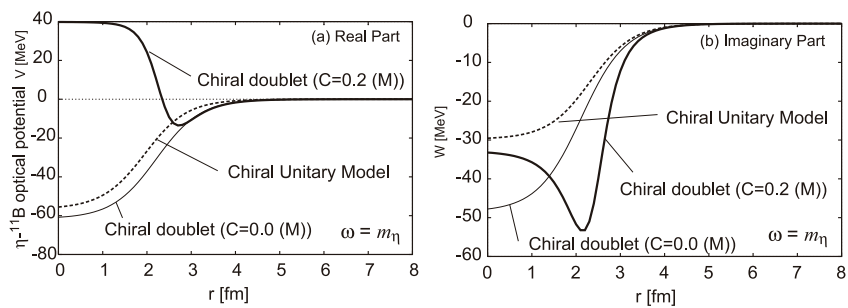}
}
\caption{
The $\eta$-nucleus optical potential for the $\eta$-$^{11}$B system as
functions of the radius coordinate $r$ reported in
 Ref.~\cite{jido02}. The left and right figures 
 show the real part and the imaginary part of the $\eta$-nucleus optical
 potential, respectively.
In both figures, the solid lines show the potentials of the chiral
 doublet model with $C=0.2$ (thick line) and $C=0.0$ (thin line), and 
 dashed lines show that of the chiral unitary model, which is 
picked 
 from the results shown in Ref.\cite{PLB550}
\label{fig:Vopt}
}
\end{figure}

In Fig.~\ref{fig:Vopt}, we show the $\eta$-nucleus optical potentials
obtained by the chiral doublet model with $C=0.0$ and $C=0.2$ 
in the case of the $\eta$-$^{11}$B system \cite{jido02}. 
As shown in the figure, the $\eta$-nucleus optical potential has a curious shape
of
a repulsive core inside nucleus and an attractive
pocket in nuclear surface. 
The qualitative feature of the optical potential discussed here,
such as the appearance of the repulsive core in the case of $C=0.2$,
is independent
of type of nuclei. Thus, 
it is 
extremely interesting to confirm the existence (or non-existence) of this
curious shaped potential experimentally.

Let us move on the second approach of the chiral unitary
model~\cite{PLB550,NPA612}.
In this approach, 
the $N^{*}$ resonance is expressed as a dynamically 
generated object in the meson baryon scattering, and
one solves a coupled channel Bethe-Salpeter equation to obtain
the $\eta$-nucleon scattering amplitude.
The optical potential in medium is obtained by closing the
nucleon external lines in the $\eta N$ scattering amplitude 
and considering the in-medium effect on the
scattering amplitude, such as Pauli blocking.
Since the $N^{*}$
in the chiral unitary approach
is found to have a large component of $K\Sigma$ 
and the $\Sigma$ hyperon is free from the Pauli blocking in nuclear
medium,
very little 
mass shift of $N^{*}$ is expected in the medium
\cite{PLB362(95)23}, while the chiral doublet
model predicts the significant mass reduction with the partial restoration of chiral symmetry 
as discussed above.
The optical potential obtained 
by
the chiral unitary approach in
Ref.~\cite{PLB550} is also shown in Fig.~\ref{fig:Vopt}, and the
potential resembles
that of the chiral doublet model with $C=0.0$
due to the small medium effect.

The calculated binding energies and widths for both potentials are reported in Refs. \cite{jido02,PLB550}.  
As we can expect from the potential shapes, they show the significant differences, which are expected to be observed in the spectra of the formation reactions of $\eta-$nucleus systems.

\subsection{Green's function method}
We use the Green's function method~\cite{morimatsu85} to calculate the formation spectra of the meson$-$nucleus systems
~\cite{jido02,Ikuta:2002kr,gata06,Koike:2007iz,YamagataSekihara:2008ji,nagahiro05,Hayano:1998sy,nagahiro05.2,jido08,nagahiro09,phi}.
We briefly explain the formalism for the ($\pi,N$) reaction case as an example.
The expected spectra of the ($\pi,N$) reactions $\displaystyle{\Bigl(\frac{d^2\sigma}{d\Omega dE_N}\Bigr)}$ are evaluated from the nuclear response function $S(E)$ and the elementary cross section $\displaystyle{\Bigl(\frac{d\sigma}{d\Omega}\Bigr)^{\rm ele}}$ in the impulse approximation as, 
\begin{equation}
\label{eq:6}
\displaystyle{\Bigl(\frac{d^2\sigma}{d\Omega dE_N}\Bigr)=\Bigl(\frac{d\sigma}{d\Omega}\Bigr)^{\rm ele}\times S(E)}~~.
\end{equation}

The calculation of the nuclear response function with a meson-nucleus complex optical potential $V_{\rm opt}$ is formulated by Morimatsu and Yazaki~\cite{morimatsu85} as,  
\begin{equation}
\label{S(E)}
S(E)=-\frac{1}{\pi} {\rm Im}\sum_f \int d{\vec r}d{\vec r}\,' \tau^\dagger_f G(E; {\vec r},{\vec r}\,') \tau_f,
\end{equation}
\noindent
where the summation is taken over all possible final states.
$G(E; {\vec r},{\vec r}\,')$ is the Green's function of the produced meson interacting in the nucleus and defined as,
\begin{equation}
\label{Gfunc}
G(E; {\vec r},{\vec r}\,')=\langle\alpha |\phi({\vec r})\displaystyle{\frac{1}{E-H +i\epsilon}}\phi^+({\vec r}\,')|\alpha\rangle~~,
\end{equation}
where $\alpha$ indicates the proton hole state and $H$ indicates the Hamiltonian of the meson-nucleus system, which includes the optical potential $V_{\rm opt}$.
The amplitude $\tau_f$ denotes the transition of the incident particle ($\pi$) to the nucleon-hole and the outgoing nucleon ($N$), involving the nucleon-hole wavefunction $\psi_{j_N}$ and the distorted waves $\chi_i$ and $\chi_f$ of the projectile and ejectile.
By taking the appropriate spin sum, the amplitude $\tau_f$ can be written as,
\begin{equation}
\label{tau}
\tau_f({\vec r})=\chi_f^*({\vec r})\xi_{1/2,m_s}^*[Y_{l}^*(\hat{\vec r})\otimes \psi_{j_N}({\vec r})]_{JM}\chi_i({\vec r})~,~
\end{equation}
\noindent
with the meson angular wavefunction $Y_{l}(\hat{\vec r})$ and the spin wavefunction $\xi_{1/2,m_s}$ of the ejectile.

The
{semi-}exclusive spectra can be calculated by decomposing the response function~(\ref{S(E)}) into the escape and conversion parts: $S=S_{\rm esc}+S_{\rm con}$.
The conversion part and the escape part are known to express the contributions to the ($\pi,N$) spectra 
of the meson absorption in nucleus and the quasi-elastic meson production processes, respectively~\cite{morimatsu85}.

\subsection{Formation by ($\pi,N$) reactions}

\begin{figure}[hbt]
\epsfxsize=11cm
\centerline{
\epsfbox{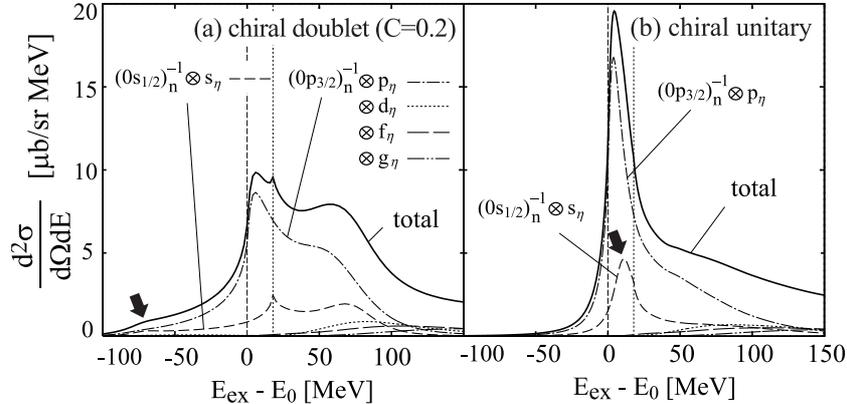}
}

\caption{
Calculated spectra of the $^{12}$C($\pi^+,p$)$^{11}$C$\otimes\eta$ reaction
 at $T_\pi=820$ MeV and the emitted proton
 angle $\theta_p=0^\circ$ 
as functions of the excited energy $E_{\rm ex}$ \cite{nagahiro09}.
$E_0$ is the $\eta$ production threshold. The $\eta$-nucleus
 interaction is calculated by using (a) the chiral doublet model with
 $C=0.2$  
and (b) the chiral unitary model.
The thick solid lines show the total spectra and the dashed lines
 represent dominant subcomponents as indicated in the figures.
The neutron-hole states are indicated as $(n\ell_j)_n^{-1}$
and the $\eta$ states as $\ell_\eta$.
The solid arrow indicates the peak due to the bound state in each model.
\label{fig:820}
}
\end{figure}

We show in Fig.~\ref{fig:820} the calculated
$^{12}$C($\pi^+,p$)$^{11}$C$\otimes\eta$ cross sections for the formation
of the $\eta$-$^{11}$C system
{with} the chiral doublet model potential
for $C=0.2$
(left panel) and that of the chiral unitary
model 
(right panel) \cite{nagahiro09}.
The incident pion kinetic energy $T_\pi$ is $820$ MeV corresponding to
the recoilless {condition} at the $\eta$ threshold. 
{
The horizontal axis indicates the excitation energy $E_{\rm ex}$ defined
as,
\begin{equation}
E_{\rm ex}=m_\eta-B_\eta+(S_n(j_n)-S_n({\rm ground}))
\end{equation}
where $B_\eta$ is the $\eta$ binding energy and $S_n(j_n)$ the neutron
separation energy from the {neutron} single-particle
level $j_n$. $S_n({\rm ground})$ indicates the separation energy from
the neutron level 
corresponding to the ground state of the daughter nucleus and
$E_0=m_\eta$.
Hence $E_{\rm ex}-E_0=0$ corresponds to the $\eta$ production threshold
with the ground state daughter nucleus.
}
In the figure, we show the total spectra {by} the
solid line and the  
contributions from {dominant} subcomponents
{by} the
dashed lines, separately. 
{We take into account the difference of the separation energy
$S_n(j_n)-S_n({\rm ground})=18$ MeV for a subcomponent accompanied by
a $(0s_{1/2})_n^{-1}$ hole-state.}
{In such a case,} the $\eta$
meson production threshold appears at $E_{\rm
ex}-E_0=18$~MeV as indicated in Fig.~\ref{fig:820} by the vertical 
dotted line.
The Fig.~\ref{fig:820} shows that the spectra are dominated by two contributions,
$(0s_{1/2})_n^{-1}\otimes s_\eta$ and
$(0p_{3/2})_n^{-1}\otimes p_\eta$, since the final states
with the total spin $J\sim 0$ are largely enhanced under the recoil-free
kinematics.

Let us see the spectra around the threshold; $-50$ MeV 
$\lesssim E_{\rm ex}-E_0 \lesssim 50$ MeV.
The spectra in this energy region were already shown in the case
of the (d,$^3$He) and ($\gamma,p$) reactions in
Refs.~\cite{jido02,nagahiro05.2}. 
The present work confirms that the spectral shape is
very similar 
with the previous calculations showing that the structure of the formation 
spectra is not sensitive to the reaction mechanism. As already discussed 
in detail in Refs.~\cite{jido02,nagahiro05.2},
the spectra {with} the ($\pi^+,p$) reaction
around the $\eta$ production  
threshold show that the repulsive nature of the 
optical potential in the chiral doublet model shifts the spectra into
the higher energy region, whereas the spectra obtained in the chiral unitary model
is shifted into the lower energy region as a result of its attractive
potential. 

{We conclude that the difference between the expected
spectra with two chiral models seems to be visible in the
($\pi^+,p$) reaction as well as the ($\gamma,p$) reaction in spite of
the distortion effect for the injected particle $\pi$.
}
The systematic results and the detail discussions are found in Ref. \cite{nagahiro09}.

\subsection{Formation by $d+d$ reactions - Preliminary results -}

The $\eta-$nucleus optical potential has large imaginary part in general, 
which makes the decay widths of the possible bound states large and makes the observation of distinct peak structures in the spectrum difficult. Thus, it is interesting to consider the systems of $\eta$ in light nuclei  because we can expect to avoid the additional difficulties due to overlaps of many subcomponents to deduce the physical information from the observed spectrum.  

In this subsection, we consider the formation reaction of the $\eta + \alpha$ system proposed in Ref. \cite{Moskal}.  
We have evaluated the formation rate of the $\eta + \alpha$ system using the Green's function method.  We have paid the special 
attention to calculate both cross sections of the $\eta$ production $d+d \rightarrow \alpha +\eta$ reaction and the bound state formation $d+d \rightarrow (\alpha \eta)_{\rm bound}$  using the same model simultaneously. 
This is important to reduce the uncertainties of the theoretical calculation due to the nuclear transition form factor in the very high momentum transfer reaction $(\sim 1 {\rm GeV/c})$ by using the  $d+d \rightarrow \alpha +\eta$ data near threshold \cite{ddeta}.  We have developed a simple theoretical model for the reaction, where we include one parameter to control the nuclear transition form factor $(d+d \rightarrow \alpha)$ with $\eta$ production in addition to the $\eta-$nucleus optical potential strength. The details of the theoretical model will be described in Ref. \cite{dd}.  

\begin{figure}[hbt]
\epsfxsize=11cm
\centerline{
\epsfbox{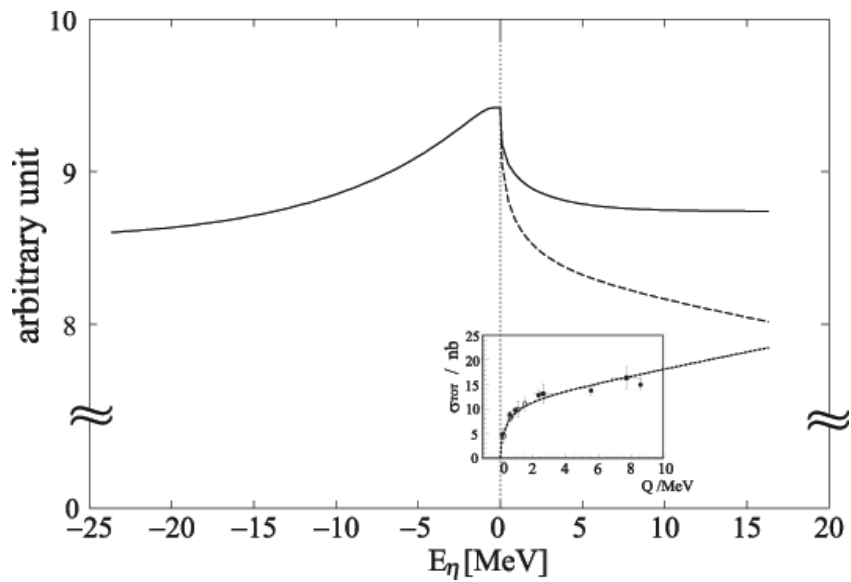}
}

\caption{
Calculated spectra of the $d+d \rightarrow \alpha +\eta$ reaction are plotted as functions of the kinetic energy of $\eta$.  The negative energy region corresponds to the formation of (subthreshold) $\eta + \alpha$ bound system.  The $\eta$ production threshold is shown as a vertical dotted line.  Solid and dashed curves  show the total spectrum and conversion part, respectively.  The escape part is also shown as a dotted curve and compared with experimental data above threshold \cite{ddeta}
\label{total_fig_50_40}
}
\end{figure}

We show the preliminary results in Fig. \ref{total_fig_50_40} where we have assumed the depth of the $\eta-$nucleus optical potential as $V_\eta (0)= (- 50 - 40i)$ $ {\rm MeV}$ at nuclear center, which are similar with those of chiral unitary model.  
The absolute value of the cross section is scaled to fit the data \cite{ddeta}.  
We found that the shape of the $d+d \rightarrow \alpha +\eta$ data were reproduced reasonably well by the escape part of the spectrum.  The conversion part corresponding to the contribution of the $\eta$ absorption process by nucleus is found to have peak structure at the threshold on the large structureless background. The results of this model will provide useful estimates for the planned experiment \cite{Moskal} where the particle pair emissions from $\eta$ absorption will be measured.   

The calculated spectrum shape is found to be sensitive to the optical potential strengths and to the (non-)existence of the bound states.  Thus, we think the experimental data will provide new information on the $\eta-$nucleus interaction.   The systematic numerical results will be reported in Ref \cite{dd}. 

\subsection{$\eta ' (958)-$Nucleus systems}

As for the formation of other meson$-$nucleus systems, we mention here another research activity.  
The $\eta ' (958)$ - nucleus systems are very interesting because the origin of its large mass is believed to be the consequences of the $U_A(1)$ anomaly effects.   
Structure and formation reaction of the $\eta ' (958)$ - nucleus systems were studied in Ref. \cite{nagahiro05} for the first time, and the formation of this system by ($\pi,N)$ reactions are also reported in Ref \cite{NFQCDdeko}, recently.  We can expect to have important information on the anomaly effects at finite density, which has not been known at all so far,   by studying the $\eta ' (958)$ - nucleus systems.

\section{Conclusion}
The mesic atoms and mesic nuclei are very interesting and fruitful objects to study the aspects of the strong interaction symmetries at finite density.  We briefly explain the recent research activities on $\eta$ mesic nuclei.   

We think that the theoretical results reported here will help to consider future experiments and to obtain new information on meson and/or  baryon resonance such as $N^*$ properties in nucleus from data. 

\section{Acknowledgement}
S. H. thanks P. Moskal for the invitation to this stimulating meeting and the hospitality during the stay in Krakow.

\end{document}